\documentclass{PoS}

\title{One-loop corrections to multiscale effective vertices in the EFT for Multi-Regge processes in QCD}

\ShortTitle{One-loop corrections in the High-Energy EFT}

\author{\speaker{Maxim Nefedov}
        \thanks{Work supported in part by the Foundation for the Advancement of Theoretical Physics and Mathematics BASIS, grant No. 18-1-1-30-1}\\
        Samara National Research University,\\
        II Institute for Theoretical Physics, Hamburg University\\
        E-mail: \email{nefedovma@gmail.com}}

\abstract{The computation of one-loop corrections to the $\gamma^\star Q_+ q$ and $gR_+g$ effective vertices in the framework of gauge-invariant effective theory for Multi-Regge processes in QCD is reviewed. Due to consistent implementation of the ``tilted Wilson line'' regularization for rapidity divergences, the gauge-invariance has been preserved at all stages of calculation independently on the rapidity regulator and cancellation of the power-like dependence on the regularization variable is traced. Only single-logarithmic rapidity divergence is left in the final result.}

\FullConference{XXVII International Workshop on Deep Inelastic Scattering and Related Subjects\\ Torino, Italy, 8-12 April 2019}

\begin{document}

\section{Introduction}

  In the Multi-Regge Kinematics (MRK) for the $2\to 2+n$ partonic scattering in QCD, the final-state partons can be grouped into clusters w.r.t. their rapidity. Different clusters are highly-separated in rapidity from each-other, so that the typical $t$-channel momentum transfer is much smaller than the invariant mass of any pair of final-state clusters. At leading power in $t/s$, higher-order QCD corrections to such amplitudes are enhanced by high-energy logarithms $\log s/(-t)$. The Gauge-Invariant Effective Field Theory (EFT) for Multi-Regge processes in QCD~\cite{Lipatov95, LipatovVyazovsky} has been introduced as a systematic tool for computation of asymptotics of QCD scattering amplitudes in the Multi-Regge limit in the Leading Logarithmic Approximation and beyond. The Hermitian version of this EFT~\cite{Lipatov_rev, Herm_action} contains all corrections, restoring the unitarity of High-Energy scattering and therefore provides a framework for studies of High-Energy QCD and gluon saturation phenomena, alternative to the Balitsky-JIMWLK or Color-Glass Condensate pictures, see Refs.~\cite{Hentsch_JIMWLK, Bond_Hier} for the recent work in this direction. 
  
  In the High-Energy EFT~\cite{Lipatov95, LipatovVyazovsky}, different rapidity-clusters of final-state particles are produced by different gauge-invariant subamplitudes -- {\it effective vertices}. This effective vertices are connected by $t$-channel exchanges of Reggeized gluons ($R_{\pm}$) and Reggeized quarks ($Q_{\pm}$), collectively named as Reggeons -- gauge-invariant degrees of freedom of the High-Energy QCD. Eventually, it should be possible to integrate-out physical quarks and gluons, order-by-order in $\alpha_s$, and formulate the high-energy limit of QCD entirely in terms of Reggeons -- {\it Reggeon Field Theory}, see e.g.~\cite{Bond_Hier, Lipatov_RFT, Bond_RFT}. Calculation of the one-loop corrections to different effective vertices is a major task in development of this formalism. 

  The main technical difficulty in the Higher-Order calculations in High-Energy EFT is the appearance of {\it Rapidity divergences} in loop and phase-space integrals. These divergences arise due to the presence of ``Eikonal'' denominators $1/l^{\pm}$ in the induced vertices of interactions of Reggeons with ordinary (Yang-Mills) partons, taken together with kinematical constraints following from MRK. See Sec. 2 of Ref.~\cite{Nefedov_loop} for the analysis of the conditions of appearance of rapidity divergences at one loop. At present, many calculations~\cite{Nefedov_loop, HAsV, MN_VS_gaQq, HAsV2l} in the High-Energy EFT has been done with the use of a variant of ``tilted Wilson line''  regularization, where the direction vectors ($n_{\mu}^{\pm}$) of Wilson lines in the definition of Reggeon-parton interactions are slightly shifted from the light-cone:
\begin{equation}
n^\pm_{\mu} \to \tilde{n}^{\pm}_\mu=n^{\pm}_\mu+r\cdot n^{\mp}_\mu,\ \ \frac{1}{l_\pm} \to \frac{1}{\tilde{l}_{\pm}}=\frac{1}{l_{\pm}+r\cdot l_{\mp}},\label{Eq:r-reg-def}
\end{equation}    
where $0<r\ll 1$ is the regularization variable. In Ref.~\cite{Nefedov_loop} we have observed, that to keep the $Rg$-interaction gauge-invariant for $r\neq 0$ one also have to modify the usual MRK kinematic constraint, stating that four-momentum $q_1$ of $R_{+}$-Reggeon has only one nonzero light-cone component $q^{+}_{1}$ and transverse momentum ${\bf q}_{T1}$. The kinematic constraint for Reggeon $R_{+}$, consistent with gauge-invariance at $r\neq 0$ is
\begin{equation}
\tilde{q}^{-}_{1}=q^{-}_{1}+r\cdot q^{+}_{1}=0. \label{Eq:mod-MRK}
\end{equation}  
  For Reggeized quarks, such modification is not strictly necessary, but it turns out, that many scalar integrals actually simplify in the kinematics (\ref{Eq:mod-MRK}), so we prefer to keep it both for Reggeized gluons and quarks.

  In the present contribution we will discuss two examples of one-loop corrections to Reggeon-Particle-Particle effective vertices: $\gamma^\star Q_+ q$ and $gR_+g$. The first one involves an off-shell photon ($\gamma^\star$), so that the vertex has two scales of virtuality: virtuality of the photon $q^2=-Q^2<0$ and of the Reggeized quark $q_1^2=-t_1<0$. More details concerning this example can be found in our Ref.~\cite{Nefedov_loop}. The second example already has been considered in Ref.~\cite{HAsV}, however in this reference part of diagrams has been but to zero by the gauge choice for external gluons and therefore gauge-invariance of amplitude and cancellation of power-like dependence on the rapidity-regulator $r$ has not been verified. We fill this gap in the present contribution. 

  Our paper has the following structure: In the Sec.~\ref{sec:Ints} integrals appearing in our calculation are listed and we comment on their rapidity divergences. Explicit expressions for this integrals are provided in Ref.~\cite{Nefedov_loop}. In Sec.~\ref{sec:Verts} we review the calculations for above-mentioned examples and in the Sec.~\ref{sec:Concl} we summarize our conclusions.   

\section{One-loop rapidity-divergent integrals}
\label{sec:Ints}

  It is convenient to categorize one-loop integrals appearing in our calculations according to the type of their dependence on the rapidity-regulator variable $r$. Then the simplest integrals containing only one quadratic and one or two linear propagators turn out to be the most singular ones. Integrals:
\[
A_{[-]}(p)=\int\frac{[d^d l]}{(p+l)^2 [\tilde{l}^-]},\ A_{[--]}(p)=\int\frac{[d^d l]}{l^2 [\tilde{l}^-][\tilde{l}^- - \tilde{p}^-]},
\]
where $[d^d l]={(\mu^2)^\epsilon d^d l}/(i\pi^{d/2} r_\Gamma)$, $d=4-2\epsilon$, $r_\Gamma=\Gamma^2(1-\epsilon)\Gamma(1+\epsilon)/\Gamma(1-2\epsilon)$ and $1/[X]$ denotes the PV-prescription for the light-cone denominator: $1/[X]=(1/(X+i0)+1/(X-i0))/2$, are related with each-other as:
\begin{equation}
A_{[--]}(p)=\frac{1}{\tilde{p}_-} A_{[-]}(p), \label{Eq:rel-Amm}
\end{equation}
and both are proportional to $r^{-1+\epsilon}$. 

  Integrals 
\[
B_{[-]}(p)=\int\frac{[d^d l]}{l^2 (p+l)^2 [\tilde{l}^-]},\ \  B_{[--]}(p)=\int\frac{[d^d l]}{l^2 (p+l)^2 [\tilde{l}^-] [\tilde{l}^-+\tilde{p}^-]}, 
\]
are related as:
\begin{equation}
B_{[--]}(p)=\frac{2}{\tilde{p}_-} B_{[-]}(p), \label{Eq:rel-Bmm}
\end{equation}
and contain the $r^{\epsilon}$-dependence on the rapidity regulator for $p^2=0$ and also term $\propto r^{-\epsilon}$ appears for $p^2\neq 0$. These power-like terms come together with $1/\epsilon^2$-factors and could lead to $\log^2 r$-dependence after expansion in $\epsilon$, which would contradict to Reggeization of gluon and quark. Cancellation of this terms happens between different diagrams and hence is a nontrivial dynamical property of QCD. 

  The integral
\[
B_{[+-]}(p)=\int\frac{[d^d l]}{l^2 (p+l)^2 [\tilde{l}^+] [\tilde{l}^-]},
\]
  contributes to one-loop correction to propagators of Reggeized gluon and quark and it contains only logarithmic rapidity-divergence $\sim \log r$, related with Reggeization. Similar single-logarithmic divergence is present in a ``triangle'' integral:
\[
C_{[-]}(-q_1^2, -q^2, q^-)=\int\frac{[d^d l]}{l^2 (q_1+l)^2 (q_1+q+l)^2 [\tilde{l}^-]},
\]
which has been computed for the case $q^2=0$ in Ref.~\cite{HAsV} and for the case $q^2\neq 0$ in the Ref.~\cite{Nefedov_loop}. For the $q^2\neq 0$ case, the term $\propto r^{-\epsilon}$ also appears in the integral $C_{[-]}$. These are all scalar integrals necessary for the calculation of one-loop corrections to Particle-Particle-Reggeon effective vertices.

\section{One-loop effective vertices}
\label{sec:Verts}

\begin{figure}[t]
\begin{center}
\includegraphics[width=0.45\textwidth]{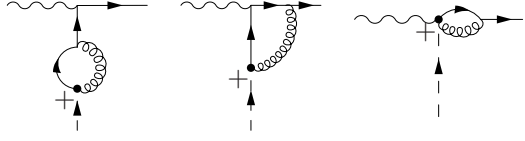}
\end{center}
\caption{ Diagrams contributing to the $\gamma^\star Qq$-vertex at one loop. Dashed line with an arrow -- Reggeized quark.}\label{fig:gaQq-diags}
\end{figure}

\begin{figure}[t]
\begin{center}
\includegraphics[width=0.5\textwidth]{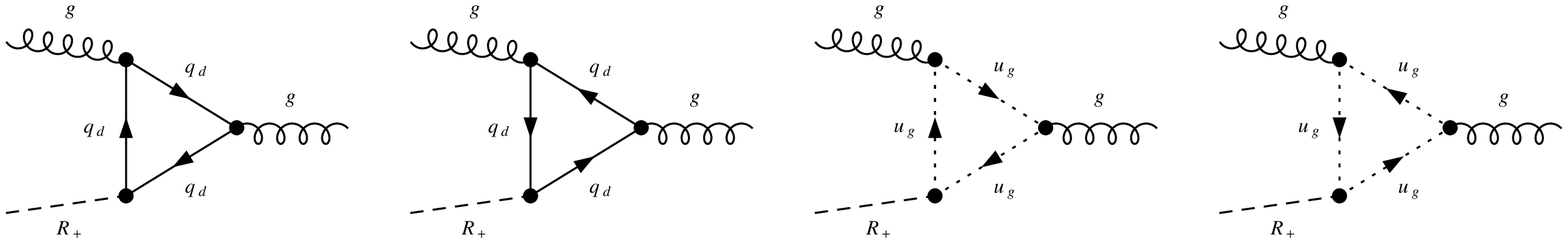}\\
\includegraphics[width=0.5\textwidth]{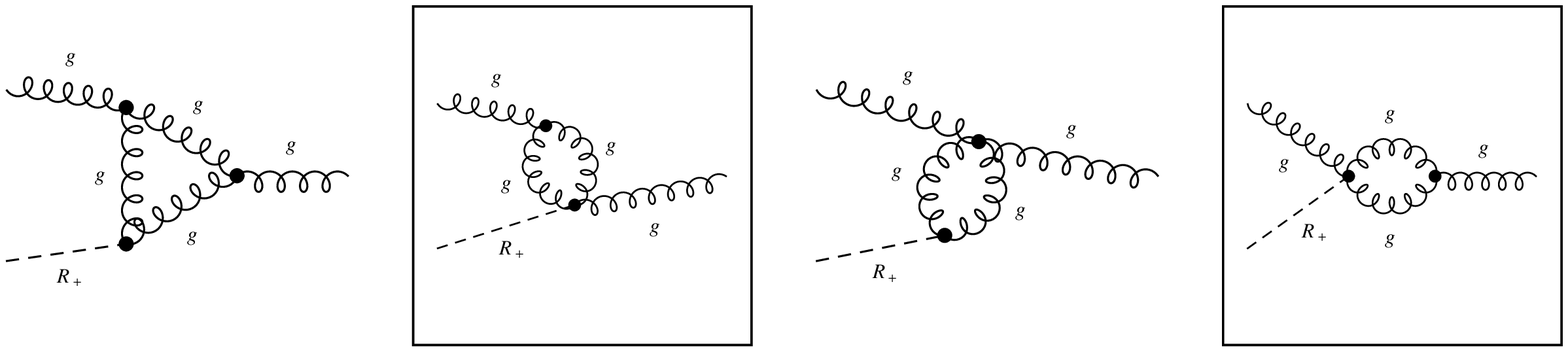}\\
\parbox{0.43\textwidth}{\includegraphics[width=0.4\textwidth]{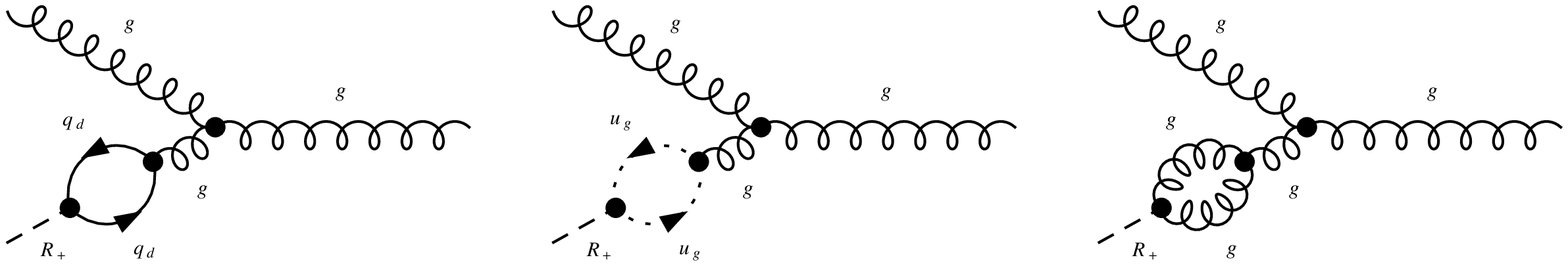}}\parbox{0.1\textwidth}{\includegraphics[width=0.1\textwidth]{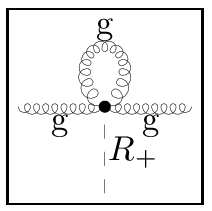}}\\
\end{center}
\caption{ Diagrams contributing to the $gR_+g$-vertex at one loop. Dashed line -- Reggeized gluon, dotted line -- Faddeev-Popov ghost.  }\label{fig:gRg-diags}
\end{figure}

  The set of EFT Feynman diagrams, contributing to the one-loop correction to $\gamma^\star Q_+ q$-effective vertex is shown in the Fig.~\ref{fig:gaQq-diags}. To compute them, we perform the tensor reduction procedure, similar to the standard one. However since now some integrals contain Eikonal denominators, depending on the vector $\tilde{n}^-_\mu$, one should include this vector to the ansatz for the tensor structure. The result~\cite{Nefedov_loop}:
\[
\Gamma_{+\mu}^{(1)}(q_1,q)=iee_q\cdot \bar{u}(q+q_1)\left[ C[\Gamma]\cdot\Gamma_{+\mu}^{(0)}(q_1,q)+C[\Delta^{(1)}]\cdot\Delta_{+\mu}^{(1)}(q_1,q)+C[\Delta^{(2)}]\cdot\Delta_{+\mu}^{(2)}(q_1,q) \right],
\] 
can be expressed in terms of three gauge-invariant Lorentz structures:
\begin{eqnarray*}
& \Gamma_{+\mu}^{(0)}(q_1,q)=\gamma_\mu + \frac{\hat{q}_1 n_\mu^-}{2 q^-},\  \Delta_{+\mu}^{(1)}(q_1,q)=\frac{\hat{q}}{q_-} \left( n_\mu^- - \frac{2(q_1)_\mu}{q_1^+} \right), \ 
\Delta^{(2)}_{+\mu}(q_1,q)=\frac{\hat{q}}{q_-} \left( n_\mu^- - \frac{q_\mu}{q^+} \right), 
\end{eqnarray*} 
where $\Gamma_{+\mu}^{(0)}$ is the Fadin-Sherman scattering vertex and coefficients are the following:
\begin{eqnarray}
  C[\Gamma]&=& -\frac{\bar{\alpha}_s C_F}{4\pi} \frac{1}{2} \left\{ \frac{[(d-8)Q^2+(d-6)t_1]B(t_1)-2(d-7)Q^2 B(Q^2)}{Q^2-t_1} \right. \nonumber \\
&& \left. -2\left[ (Q^2-t_1)C(t_1,Q^2)- q_- \left( t_1 C_{[-]}(t_1,Q^2,q_-)+(B_{[-]}(q)-B_{[-]}(q+q_1)) \right) \right] \right\}, \label{Eq:C-Gm} \\
 C[\Delta^{(1)}]&=& -\frac{\bar{\alpha}_s C_F}{4\pi} \frac{(Q^2+t_1) }{2(Q^2-t_1)^2}\left[ \left( (d-2)Q^2-(d-4)t_1 \right)B(t_1) -2Q^2 B(Q^2) \right] ,\label{Eq:C-D1} \\
 C[\Delta^{(2)}]&=& -\frac{\bar{\alpha}_s C_F}{4\pi} \frac{Q^2}{(Q^2-t_1)^2}\left[ \left( (d-6)t_1-(d-8)Q^2 \right)B(Q^2) +2(t_1-2Q^2) B(t_1) \right] ,\label{Eq:C-D2}
\end{eqnarray}  
were $\bar{\alpha}_s=\mu^{-2\epsilon} g_s^2 r_\Gamma/(4\pi)^{1-\epsilon}$ is the dimensionless strong-coupling constant, $B(t_1)$ and $C(t_1,Q^2)$ are the usual one-loop scalar ``bubble'' and ``triangle'' integrals~\cite{Ellis_1-loop}.  We observe, that integrals $A_{[-]}(q)$ appearing in the expansion of the second and third diagrams in the Fig.~\ref{fig:gaQq-diags} cancel-away. Also the terms $\propto r^{\pm \epsilon}$ cancel between integrals $B_{[-]}(q)$, $B_{[-]}(q+q_1)$ and $C_{[-]}(t_1,Q^2,q_-)$ in Eq.~(\ref{Eq:C-Gm}), so that only single-logarithmic rapidity divergence is left. In Ref.~\cite{Nefedov_loop} we have checked, that this rapidity divergence cancels in the single-Reggeon exchange contribution to the $\gamma^\star + \gamma \to q+\bar{q}$-amplitude at one loop and EFT result agrees with MRK limit of one-loop QCD amplitude. 

  Diagrams contributing to the one-loop correction to $gR_+g$-vertex with on-shell external Yang-Mills gluons $g$ with helicities $\lambda_1$ and $\lambda_2$ and momenta $q$ and $q+q_1$ are shown in the Fig.~\ref{fig:gRg-diags}. This one-loop correction can be decomposed as:
\[
   \gamma_{\lambda_1 + \lambda_2}^{abc, (1)} = ig_s f^{abc}\cdot \epsilon^{\mu}(q,\lambda_1) (\epsilon^{*}(q+q_1,\lambda_2))^\nu \left[ C\left[ \gamma_+^{(0)} \right] \cdot \gamma_{\mu,+,\nu}^{(0)}  + C\left[ \delta_+ \right] \cdot \delta_{\mu, +, \nu}   \right],
\] 
 where the helicity-conserving (Lipatov's) and helicity-flip Lorentz structures are:
\begin{eqnarray*}
& \gamma_{\mu, +, \nu}^{(0)} =  2 q_- g_{\mu\nu} + 2 n_{-\mu} q_{1\nu} - 2 n_{-\nu} q_{1\mu} +\frac{t_1 n_{-\mu} n_{-\nu} }{q_-} ,\  \delta_{\mu, +, \nu} =  2q_- \left[ g_{\mu\nu} + \frac{2q_{1\mu} q_{1\nu}}{t_1} \right],
 \end{eqnarray*}
while coefficients in front of them read:
\begin{eqnarray}
  C\left[ \gamma_+^{(0)} \right] &=& -\frac{\bar{\alpha}_s C_A}{4\pi} \left[ q_- t_1 C_{[-]}(t_1,0,q_-) + B(t_1) \right], \label{Eq:hel-cons} \\
  C\left[ \delta_+ \right] &=& \frac{\bar{\alpha}_s}{4\pi} \frac{(d-4) B(t_1)}{2(d-1)(d-2)} (2n_F-(d-2)C_A). \label{Eq:hel-flip}
\end{eqnarray}

  Eqns.~(\ref{Eq:hel-cons}) and (\ref{Eq:hel-flip}) coincide with the results of Ref.~\cite{HAsV}, however in the calculations in this paper, the diagrams framed in the Fig.~\ref{fig:gRg-diags} where nullified by the gauge-choice for external gluons. We take them into account, and hence we can check the Slavnov-Taylor identities and trace-out the cancellation of power-like dependence on the regulator $r$. Modified kinematical constraint (\ref{Eq:mod-MRK}) guarantees the gauge-invariance of amplitude in all orders in $r$, and we observe, that contributions of integrals $A_{[-]}(q)$ and $B_{[-]}(q)$ cancel in the $O(r^1)$ and $O(r^0)$ respectively, while in higher orders in $r$ (which we eventually drop), coefficients in front of this integrals are gauge-invariant, which serves as a useful cross-check of the calculation. Cancellation of contributions of this integrals happens between different diagrams and essentially relies on relations (\ref{Eq:rel-Amm}) and (\ref{Eq:rel-Bmm}), while integral $B_{[-]}(q_1)=0$ due to the constraint (\ref{Eq:mod-MRK}). Therefore all power-like dependence on the rapidity-regulator cancels in the leading power in $r$  and we are again left with single-logarithmic rapidity divergence related with gluon Reggeization. 

\section{Conclusions and discussion}
\label{sec:Concl}

  In the present contribution we have reviewed the structure of rapidity divergences in the one-loop integrals contributing to the one-loop corrections to Particle-Particle-Reggeon effective vertices in the gauge-invariant EFT for Multi-Regge processes in QCD~\cite{Lipatov95, LipatovVyazovsky} and illustrated their application on two examples of such vertices: $\gamma^\star Q_+ q$ and $gR_+g$. The first one contains two scales of virtuality: squared transverse momentum of Reggeized quark $t_1$ and virtuality of the photon $Q^2$, and new Lorentz structure $\Delta_{+\mu}^{(2)}$ appears in the answer for $Q^2\neq 0$. Cancellation of power-like dependence on rapidity regularization parameter $r$ is observed in both cases, so that only single-logarithmic rapidity divergence is left in the end.

\end{document}